\documentclass[12pt]{article}
\input{amssym.def}
\input{amssym}
\usepackage[dvips]{color}
\usepackage{epsfig}
\usepackage{hyperref}
\normalbaselines
\makeatletter
\@addtoreset{equation}{section}
\makeatother

\def\be{\begin{equation}}
\def\ee{\end{equation}}
\def\ba{\begin{eqnarray}}
\def\ea{\end{eqnarray}}

\newcommand{\rf}[1]{(\ref{#1})}

\def\bra#1{\langle #1|}
\def\ket#1{|#1\rangle}

\begin{document}

\title
{Nonlocal asymmetric exclusion process on a ring and \\
conformal invariance }

\author{Francisco C. Alcaraz$^1$   and 
Vladimir Rittenberg$^{2}$
\\[5mm] {\small\it
$^1$Instituto de F\'{\i}sica de S\~{a}o Carlos, Universidade de S\~{a}o Paulo, Caixa Postal 369, }\\
{\small\it 13560-590, S\~{a}o Carlos, SP, Brazil}\\
{\small\it$^{2}$Physikalisches Institut, Universit\"at Bonn,
  Nussallee 12, 53115 Bonn, Germany}}
\date{\today}
\maketitle
\footnotetext[1]{\tt alcaraz@if.sc.usp.br}
\footnotetext[2]{\tt vladimir@th.physik.uni-bonn.de}

\begin{abstract}
 We present a one-dimensional nonlocal hopping model with exclusion on a 
ring. The model is related to the Raise and Peel growth model. A 
nonnegative parameter $u$ controls the ratio of the local backwards and 
nonlocal forwards hopping rates. The phase diagram and consequently the 
values of the current, depend on $u$ and the density of particles. In the 
special case of half-filling and $u = 1$ the system is conformal 
invariant and
an exact value of the current for any size $L$ of the system is 
conjectured and
checked for large  lattice sizes in Monte Carlo simulations. For $u > 1$
 the current has a non-analytic dependence on the density 
when the latter approaches the half-filling value. 
\end{abstract}

\section{ Introduction} \label{sect1}

  One-dimensional lattice stochastic models have received a
lot of attention during the last decades. Not only it is easier to reveal their properties through
analytical and numerical methods but they also have interesting applications
which include traffic \cite{NSM} granular gases \cite{TOR}, 
the ribosomal motion of
mRNA \cite{SZL} bio-polymerization on nucleic acid templates \cite{PFF} 
and statistics
of DNA alignment \cite{BWH}.

 The most researched model of this kind is the asymmetric exclusion process
(ASEP) \cite{DEHP}. One takes a one-dimensional chain with $L$ sites covered with
particles and vacancies and   consider periodic or open boundary
conditions where one has sources and sinks. Using sequential updating, one
chooses one particle, if the neighboring sites are empty with a rate $u$
the particle hops to the left (anticlockwise) or with a rate 1 to the
right (clockwise){\footnote { 
 This notation for the rates is kept through this paper.}.
 The model is obviously local. The stationary state
and more importantly, the dynamics of the model is mostly understood. The
critical domain is in the KPZ \cite{KPZ} universality class with a dynamic critical
exponent $z = 3/2$. The mathematics of the model is very rich and is related to
random matrix models (see \cite{PLF}  and references therein),  
and to the Bethe 
ansatz
for quantum chains with non diagonal boundary conditions \cite{JDG}.

 Among the extensions of ASEP, some consider nonlocal hopping. The model
examined in \cite{SNU} is an obvious generalization of ASEP, hoppings don't take
place only on empty neighboring sites but also on empty sites at a distance $k$
with a probability $P(k) \sim  k^{-(1+a)}$. The dynamic critical exponent
changes accordingly $z = \mbox{min} (a,3/2)$. Another generalization \cite{PND} 
 consists
in taking hoppings of two kinds: with a probability $p$ the particle hops
all the way forward to the vacant site immediately behind the 
closest  particle and with a probability $(1-p)$ on the next neighboring site, provide its empty. This
model stays in the KPZ universality class. Another way to introduce 
nonlocality in the model is to consider hopping through avalanches. 
It was shown in \cite{PRI} that depending on the density one can be 
in the KPZ universality class or in the diffusion ($z=2$) universality 
class. 

 A model much closer to the one we are going to present is the PushASEP
model \cite{BFE}. In this model, the particle hops locally to the left
(anticlockwise) on the neighboring site, provide it is empty, 
 with a rate $u$ and
non locally to the right (clockwise) on the next vacant site. The beauty of
the model is that one can do analytic calculations. When we are going
to compare this model with ours, two exact results obtained for PushASEP
are going to be relevant: the dynamic critical exponent is again $z = 3/2$
and the current in a ring is a smooth function of $u$ and the density of
particles $\rho$.

 In the model we are going to describe and for which we use the acronym
NASEP, like in PushASEP, a particle hops locally to the left with a rate $u$ and
nonlocally to the right with a rate equal to 1. The crucial difference between
the two models is the nonlocal hopping to the right. As we are going to
explain, in NASEP the distance of the vacant site where the particle hops
depends on the number of particles and vacancies encountered in the path.
It turns out that this "innocent" change will have dramatic consequences
We are going to show that if we consider the  ring geometry, one has a
phase diagram depending on both $u$ and $\rho$ presenting gapless and gapped
phases with different properties.

 Before presenting our model,  in Section 2, we would like
to explain how we "discovered" it (some not mathematical oriented readers
might prefer to skip this part of the introduction and the Section 3 of
the paper). It all started with the Raise and Peel
model (RPM) \cite{GNPR,CAR} of a fluctuating interface for an open system. 
This is
a stochastic model where, in continuum time, the evolution rules are
prescribed by a Hamiltonian which is a sum of generators of a
Temperley-Lieb algebra (TL) where the parameter is fixed such that the
generators form a semigroup. The vector space (configuration space) in
which the Hamiltonian acts is given by RSOS (Dyck) paths. These paths represent
the interface between a fluid deposited on a substrate and a rarefied
gas. This model has two important features: firstly, the stationary state
has wonderful combinatoric properties \cite{RAST,CAS} 
which make possible to make
conjectures about the size dependence of several observables \cite{MNG}; secondly,
in the finite-size scaling limit, the spectrum of the Hamiltonian is known and is
given by characters of the Virasoro algebra. This is possible since one
has conformal invariance and the dynamical critical exponent $z = 1$.
Like in the previously discussed models, the model is local for left
"moves" and nonlocal for right "moves". This is the $u = 1$ case discussed
above. The model was also extended \cite{GNPR,CAR} for $u \neq 1$. If $u > 1$ the local
hopping to the left is enhance whereas for $u < 1$ the nonlocal hopping to
the right is enhanced. New physics shows up
at the price of losing integrability and mathematical beauty.  As we will
show in Section 2, this model can be mapped in a hopping model on a
segment without sources or sinks and with an equal number 
of particles and vacancies  (half-filling).  In its new
formulation, presented in this paper, one can naturally  consider arbitrary densities.

 We consider, for the first time, an extension of the RPM to
the periodic boundary conditions case (see Section 2). This opens the
possibility to study new physical phenomena (there are for example no currents
in an open system without sources and sinks but they might exist in the
periodic system). We start with the periodic Temperley-Lieb algebra (PTL) 
\cite{MAS} at
the semigroup point in which we use the link presentation. This gives us the
model for $u = 1$ and density of particles $\rho$ = 1/2. One obtains a stochastic model with
again  fascinating combinatorial properties of the probability distribution
function describing the stationary state \cite{RAST,CAS}. The spectrum of the
Hamiltonian can be obtained using the Bethe ansatz  and, in the
finite-size scaling limit, it can be written in terms of representations of
the Virasoro algebra since the system, like the open case, is conformal invariant (see appendix B). The model
generalizes naturally for any values of $u$ and for any densities of 
particles $\rho$.
All the algebraic considerations can be found in Section 3.

 In Appendix B, again for $u = 1$, we give the connection between our model
and the XXZ spin 1/2 quantum chain. This connection becomes relevant when
we discuss the currents.

 From now on, the text should be of interest to any reader. In Section 2
we explain the model for open and periodic boundary conditions for any
backward-forward asymmetry $u$ and any density of particles $\rho$. 
The model is
presented in two versions. The first one is in terms of particles and
vacancies, this is the NASEP version of the model. The second version is
in terms of charged particles (positive and negative), in this case in the
sequential updating we don't chose a particle like in NASEP but a bond.
The last version is simpler and has a direct connection with the
arguments presented in Section 3. If one makes the substitution positive
particle $\rightarrow$ particle, negative particle $\rightarrow$ vacancy, 
the two versions of
the model give identical results.

 In Section 4 we discuss the half-filling case. For $u < 1$ the system is
gapped, for $u = 1$ it is gapless and conformal invariant ($z = 1$), for $u > 1$
it is gapless but not conformal invariant ($z<1$ decreases continuously when $u$
increases). In the stationary state the current vanishes in the thermodynamical limit for any $u$
but its behavior, as a function of the size of the system $L$, reflects the
phase diagram. For $u < 1$, it vanishes exponentially. For $u = 1$, we find
for large values of $L$ ($L$ even) the expression
\begin{equation} \label{1.1}
J(L) = v_s C/L,                         
\end{equation}
where $v_s$ is the sound velocity and $C$ a universal constant which was
determined. The current vanishes identically if $L$ is odd. The
explanation of this phenomenon is given in Section 4 and Appendices A and B. 
We also present the  $L$ dependence of the dispersion of 
the current in the stationary state and of the time dependence of the 
current. These results were obtained using Monte Carlo simulations. For 
 $u > 1$
the current vanishes like a power $J(L) \sim 1/L^x$ where the exponent $x$, like the
exponent $z$, decreases when $u$ increases.

 The currents for other densities are discussed in Section 5. For any
values of $u$ the system is gapped and the currents finite in the
thermodynamical limit. This implies that if we let the density approach
the value 1/2, we have no phase transition if $u < 1$, a usual phase
transition if $u = 1$ and possibly a new kind of phase transition if $u > 1$. We find
indeed that for $u < 1$ and $u = 1$ the current vanishes smoothly if the
density approaches the value 1/2 
but it has a non analytical dependence on the density if $u > 1$. In order 
to understand the non-analytic behavior we had to use lattices up to 
256000 sites in our Monte Carlo simulations.

 Finally, in Section 6 we summarize the long list of unanswered questions.

\section { A model for nonlocal asymmetric exclusion processes (NASEP)}

 We present two equivalent versions of the same model. One in terms of
particles and vacancies (this is the NASEP formulation of the model), the
other one in terms of charged particles. The first version has an
obvious physical interpretation while the second is simpler and has a
transparent connection with quantum chains. We give the rules
when the hopping processes take place on a ring and in an open system.

 One takes a lattice with sites $1 \leq i \leq	 L$ and fill it with particles and
vacancies (first version of the model) or with positive and negative
particles (the second version of the model). We use sequential updating.
The continuous time evolution of the system is given by the following rules:

{\bf a) The particles-vacancies version of the model.}

1) On a given site one can have at most one particle (exclusion).

2) If on the site $i$ one has a particle and the preceding site $i-1$ is empty,
with a rate $u$ the particle hops to the left, filling the site $i-1$, and with a
rate 1 hops to the right to the empty site $i+k$. The site $k$ is chosen such that
there are an equal number of particles and vacancies in the segment ($i,
i+k$) AND the site $k+1$ is also empty. The number $k$ is the smallest number which
satisfies these conditions (see Fig.~1a). If the site $k+1$ is full,
the hopping to the right is forbidden and the particle hops only to the left
with the rate $u$ (see Fig.~1b).

3) If on the site $i-1$ one has a particle, the particle on the site $i$ can
hop only to the right to the site $k$ chosen as before. If the site $k+1$ is
empty the hopping takes place with a rate equal to $2$ (Fig.~1c). If the site
$k+1$ is full the rate is equal to 1 (see Fig.~1d)

\begin{figure}
\centering
\includegraphics[angle=0,width=0.5\textwidth] {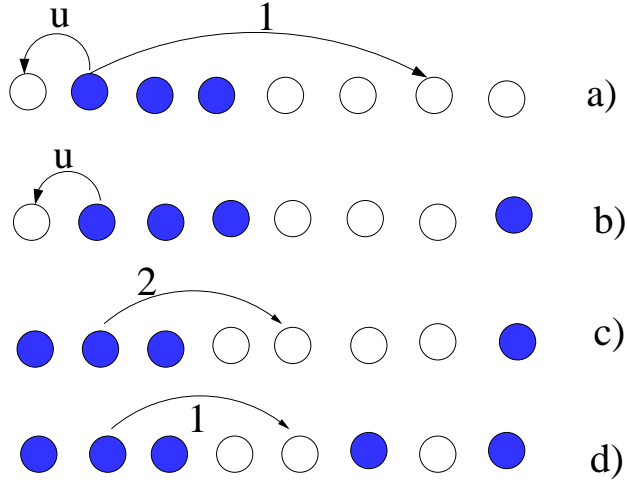}
\caption{
       a,b) Hopping rules if the particle is preceded by a
      vacancy. c,d) Hopping rules if the particle is preceded by
      another particle.}
\label{prof1}
\end{figure}
 The hopping to the right takes place by permuting a particle with a
vacancy leaving an equal number of particles and vacancies unperturbed.
 The rules conserve the number of particles and can be used as such as
a hopping model on a ring.

 For an open system with $L$ sites, in order to apply the rules, one has
to add a fictitious  site $L+1$ and assume that the sites $i=1$ and $i=L+1$ are
always occupied and the site $i=L$ is always empty.

 Notice that in the present model, the movement to the left is local, like
in ASEP, but the movement to the right is nonlocal unlike ASEP

{\bf b) The charged particles version of the model.}

 In this formulation of the model the sites are occupied by positive
particles, which correspond to the particles in the previous formulation of
the model, or by negative particles which correspond to the vacancies. The
rules for the time evolution are now given considering bonds connecting two consecutive sites. A
bond connecting two consecutive sites is indicated by $[c_{i-1}, c'_i]$, where $c$ and $c'$ indicate 
the charges
of the particles on the sites $i-1$, respectively $i$. The site $i$, occupied by a
charge $c=\pm$, is indicated by $(c)_i$:

%

\ba \label{2.1}
  &&      [(+)_{i-1},  (-)_i] \rightarrow   [(+)_{i-1},  (-)_i] \quad {\mbox{ 
stays unchanged}}\\            
  &&      [(-)_{i-1},(+)_{i}] \rightarrow   [(+)_{i-1},(-)_i] \quad \mbox{rate} \quad u \label{2.2}\\ 
  && [(+)_{i-1},(+)_i] + (-)_{i+k}   \rightarrow 
   [(+)_{i-1},(-)_{i}] + (+)_{i+k}    \quad \mbox{rate} \quad 1 \label{2.3}\\
 && (+)_{i-k} + [(-)_{i-1},(-)_i] \rightarrow  
(-)_{i-k}+[(+)_{i-1},(-)_i] 
 \quad \mbox{rate} \quad 1.\label{2.4}
\ea
 In (2.3) and (2.4) $k$ is chosen such that segment ($i,i+k$)  and 
($i-k,i-1$) contains an equal number
of positive and negative particles, $k$ being the smallest number which
satisfies this condition. 
These rules are
illustrated in Fig.~2 and 
\begin{figure}
\centering
\includegraphics[angle=0,width=0.3\textwidth] {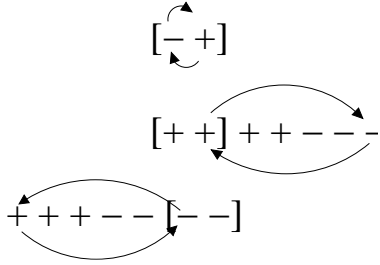}
\caption{
       Hopping rules for a bond connecting two charged
      particles.}
\label{prof2}
\end{figure}
    apply to both periodic boundary conditions as
      well as for an
      open system. As one sees in Fig.~2 the hoppings take place
      by permuting the
      end particles of a neutral domain.

 Let us stress that the main difference between the two versions of the
model is the way the sequential updating is done in a Monte Carlo
simulation. In the NASEP version one chooses randomly a particle whereas in
the charged particle version, one chooses randomly a bond.

 In the case of the ring geometry, by convention, if a particle (positive 
particle) 
 hops from a site $i$ to a  site $j$ with $j > i$, the particle moves
clockwise.

 In the next section we will show that if $u = 1$, and one takes an equal number
of particles and vacancies (equal number of positive and negative particles)
and  choose periodic boundary conditions the rules described above
coincide with those obtained from a Hamiltonian which is a sum of generators
of the periodic Temperley-Lieb algebra (TLP) \cite{MAS}. In this special case, the
system is conformal invariant. The present model is just an extension of the
TLP case to different densities of particles and to a whole range
of positive values of $u$.

      For the open system, for $u = 1$ and some special initial
      conditions, the
      rules can be derived from a Hamiltonian given by a sum of
      generators of the
      Temperley-Lieb (TL) algebra and the model coincides with the 
RPM \cite{GNPR,CAR}.

\section{Representations of the Temperley-Lieb, and periodic Temperley-Lieb 
algebras and their connections to stochastic processes with conformal
 invariance } 

In the case of an open system, the rules described in the last section
were suggested by a simple mapping of the rules of the RPM  and
extending them to more general cases. This model
was extensively studied in the past \cite{GNPR,CAR}. The RPM model is equivalent to
the one presented in Section 2 if one has $u = 1$, equal densities
of particles and vacancies and certain initial conditions. In the RPM the
time evolution of the system is given by a Hamiltonian which is a sum of
generators of the Temperley-Lieb (TL) algebra for a special value of its
parameter (see below). It is known  that in this special case one can
obtain exact results. The RPM was already studied away from the value $u = 1$. 
Using a new presentation of the TL algebra described below and used
in Section 2, we can study the evolution of the system for arbitrary initial
conditions and densities of particles. Moreover one can give a new
interpretation of the RPM using the NASEP picture instead of the interface
growth interpretation used up to now. We will shortly review the open case
below.

 The ring geometry presented in this paper is entirely new and, as it is
well known, stochastic processes on a ring have different properties than
in an open system. This makes the model interesting. In order to define
the model on a ring, we follow the same strategy as for the open system. This
implies the following steps:

a) We consider the periodic Temperley-Lieb (TLP) algebra 
 at the special point where it becomes a semigroup (the generators have the 
semigroup property, see below). The properly defined Hamiltonian written in terms of the
generators of the TLP algebra defines a stochastic process in the vector
space of monomials of the generators.

b) We use the spin representation of the algebra. In this representation
the Hamiltonian is given by an XXZ quantum chain with an anisotropy
parameter $\Delta = -1/2$ The chain has  periodic boundary conditions if the number of
sites $L$ is odd or has  twisted boundary conditions (twist angle 
$\phi = -2\pi/3$) if $L$ is even \cite{MNG}. These
quantum chains are integrable and their spectra are known in the finite-size
scaling limit, since one has conformal invariance (see appendix B). 

c) We consider the link representation of the algebra which corresponds to
the total spin $S^z = 0$ ($L$ even) and $S^z = 1/2$ or $-1/2$ ($L$ odd) 
sectors of the quantum
chains.

d) We map the link representation to a charged particles presentation of
the algebra (this is an essential step). The action of the generators in
this presentation is precisely given by the rules (2.1-2.4) of Section 2 for
$u = 1$ and an equal number of positive and negative particles. The charged
particles presentation can be mapped into a path one which is,
as we are going to see, also useful.

e) We apply the same rules for arbitrary densities and introduce the
parameter $u$ which favors the clockwise movements of the particles ($u < 1$)
or anticlockwise movements ($u > 1$).

f) We map the action of the whole Hamiltonian (not of each generator) in
the particles-vacancies presentation which gives the NASEP model.

 We present now the whole construction of the model.
 The evolution operator of a system with $L$ sites is given by
\be \label{3.1}
H = \sum_{k=1}^{L-1} (1 - e_k),
\ee
for the open system and
\be \label{3.2}
H = \sum_{k=1}^L (1 - e_k),
\ee
for the periodic system.
$e_k$ ($k = 1,2,\ldots  ,L-1$) are the generators of the Temperley-Lieb 
algebra (TL),
and $e_k$ ($k = 1,2,\ldots,L$), $e_k = e_{k+L}$ are the generators of the
periodic TL algebras at the semigroup point:
\be  \label{3.3}
e_{k\pm 1}e_ke_{k\pm 1} = e_k, \quad e_k^2 = e_k  \quad (k = 1,2,\ldots,L-1) \quad  [e_k, e_l] = 0 \quad (|k-l| >
1).
\ee 
For $L$ even the periodic TL algebra has a supplementary condition which is
going to be
discussed below.
 The representation of the Hamiltonians (3.1) and (3.2) in terms of 
Pauli matrices are given in appendix B.
 We are looking for other representations of the TL algebras which
give 
invariant subspaces (representations of ideals of the algebra). We
consider
first the open system.

{\bf a) The open system.}

 The configuration space of the TL representation we consider, has
dimension $L!/(L/2+1)([L/2]!)^2$. Since this representation   
in terms of link
patterns and  Dyck paths,  is well known 
we  take the simple example $L = 4$ to make the connection with
NASEP
described in Section 2. 
  We  fix the vector space (configuration space) in
which the
TL algebra acts. In the link representation of the TL algebra, the
four sites
are linked by non-intersecting arches (see Fig.~3a), the
representation has
dimension 2. The Dyck path representation is obtained considering the
dual
lattice ($\tilde{i}=0,1,\ldots,L$) and counting on each site how many times arches are crossed
(Fig.~3b). It
is useful to see the Dyck paths as being the border of an aggregate of
tiles on top of a substrate. The right hand side of Fig.~3b is the
substrate
and in the left side of the picture there is one tile deposited on the
substrate. 

 In the charged particles presentation one considers the slopes in the Dyck
path. Each up (down) step corresponds to a positive (negative) particle.
Notice that in both configurations one has an equal number of positive and
negative particles but not all six configurations with two positive and
negative particles are allowed. Only configurations in which on the left
of each bond there are no more negative particles than positive ones 
play a role. 
Notice that in the model described in Section 2 all six configurations
with two positive and two negative particles were considered. We will
return to this point later.

 In the particles-vacancies presentation, the positive particles are
replaced by particles and the negative one by vacancies (Fig.~3d). The
reason for using different notations for the last two vector spaces will
become apparent when we will describe the action of the Hamiltonian in
these vector spaces.

\begin{figure}
\centering
\includegraphics[angle=0,width=0.3\textwidth] {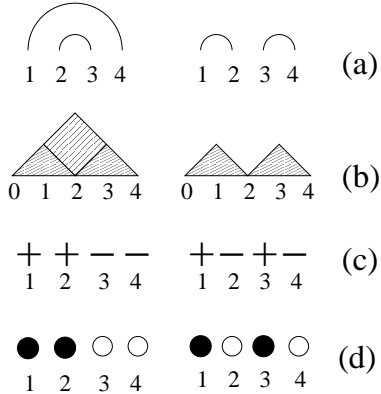}
\caption{
Different presentations of the vector space in which the
generators of
the Temperley-Lieb algebra acts for $L = 4$ sites. a) The arch
presentation,
b) The Dyck path presentation, c) The charged particles
presentation d) The
particles-vacancies presentation.}
\label{prof3}
\end{figure}
Let us compute (see Fig.~4) the action of the generators $e_1$ and $e_3$
on the
left configurations of Figs.~3a,b,c. For Fig.~4a we have used the
standard action
of the generators of the TL algebra on link patterns \cite{GNPR}. For
Fig.~4b we
have used the rules of the RPM \cite{GNPR,CAR}, for Fig.~4c we
have
used (2.3) and (2.4) with $u = 1$. The same results are obtained
using the
mappings shown in Figs.~3a,b,c. This implies that in this configuration
space,
where we have desorption only (a tile is lost in the process), 
the dynamics of the system is given by
the
Hamiltonian \rf{3.1} and the charged particles version of the model  coincide.

\begin{figure}
\centering
\includegraphics[angle=0,width=0.5\textwidth] {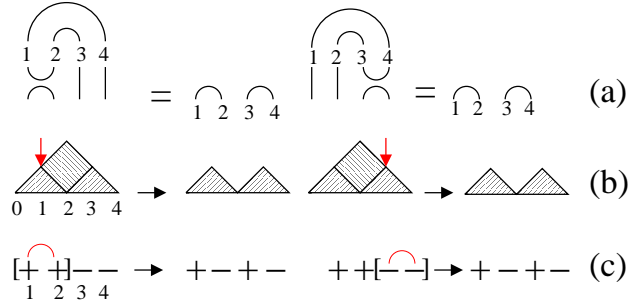}
\caption{
 The action of the generators $e_1$ and $e_3$ on the configurations
shown in
the first column in Fig.~3. a) The arch presentation, b) The path
presentation, c) the charged particle presentation. All actions
take place
with a rate equal to 1.
}
\label{prof3b}
\end{figure}
  In the particle-vacancy picture (NASEP), only the particle moves, this
implies that one has to consider the action of the sum of the
generators $e_1$
and $e_3$ and one gets a factor of 2 (see Fig.~5 ) in agreement with the
rule 3)
of Section 2 (see also Fig.~1c).
\begin{figure}
\centering
\includegraphics[angle=0,width=0.3\textwidth] {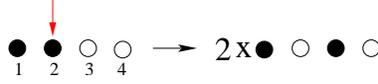}
\caption{
 The hopping of the particle on the site 2 to the site 3 with a
rate 2
corresponds to the action of the sum $e_1+e_3$ in Fig.~4.
}
\label{prof4}
\end{figure}
  What we have shown above is that for the 2 configurations in Fig.~3
the
action of the Hamiltonian \rf{3.1} 
which produces the desorption of
a tile, coincides with the dynamics given by the rules of Section 2  (Fig.~3b).

 A first generalization consists in changing the adsorption rules. This
corresponds to the hopping to the left (Fig.~1a) which is equivalent to a  
permutation \rf{2.2} in which a positive particle moves to the left and the
negative particle to the right. This move corresponds to the action of
the generator $e_2$ on the configuration shown in the right column of 
Fig~.3a
(see Fig.~6). The action of $e_2$ coincides with the rules of Section 2
(see Fig.~1a) and \rf{2.2} only if $u = 1$. If we follow the rules of Section 2 with
$u \neq 1$, the evolution of the system is not anymore related to the TL 
algebra.

\begin{figure}
\centering
\includegraphics[angle=0,width=0.3\textwidth] {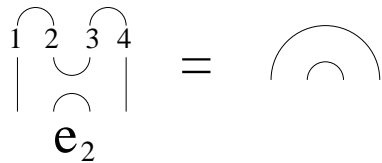}
\caption{
 In the TL algebra $e_2$ acts with a rate $u = 1$.
}
\label{prof6}
\end{figure}

 A second generalization of the model is obtained when we do not 
 restrict ourselves to the two
configurations of Fig.~3 by considering the 6 possible configurations with two
particles and two vacancies (two positive and two negative particles
respectively). Now the language of arches and Dyck paths is not useful
anymore. The model describes the movement of particles in a larger vector space
but one can show that the two configurations shown in Fig.~3 form an
invariant subspace. This implies that if the initial conditions of the
stochastic process contain only these two configurations, in the
evolution of the system the remaining 4 configurations will not show
up and therefore the stationary state coincides with the one obtained
if one considers the two configurations only.

 The discussion presented here for the 4 sites problem generalizes for any
number of sites. Moreover the models discussed in Section 2 can be used for
any densities of particles. We should mention that the inclusion of
defects \cite{PRA} in the representation of the TL algebra gives a different
dynamics than the one considered in the present model 
since the number of particles is not conserved in this case. This closes the
discussion of NASEP for an open system. In the present paper we will not
discuss the properties of the open system with an enlarged space. 
We hope to return to this
problem in the near future.

{\bf  b) The periodic system.}

 We are going now to show that the same rules which have defined NASEP for the
open system stay valid for a periodic system. For $u = 1$ and
half-filling they coincide with those obtained from the Hamiltonian \rf{3.2} in
which one uses the generators of the TLP algebra. The number of
configurations is $L!/([L/2]!)^2$. We restrict ourselves again to the $L = 4$ 
example.

\begin{figure}
\centering
\includegraphics[angle=0,width=0.5\textwidth] {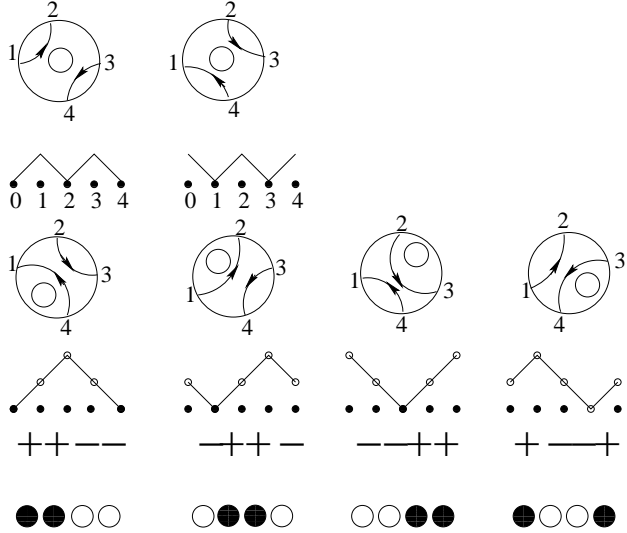}
\caption{
 The periodic system for $L = 4$. The six configurations are shown
in the
arch, path, charged particles and particles-vacancies presentations.
 }
\label{prof7}
\end{figure}

 The arch presentation of the algebra is given by oriented non intersecting
arches on a puncture disc (Fig.~7) \cite{MAS}. We have assigned charge particles to the sites
using the following rule: if an arch starts on the site $i$ and ends on a
site $j$ and does not contain the puncture, we assign a charge $(+)$ to the
site $i$ and a charge $(-)$ to the site $j$. 
If the arch contains the puncture of
the disc, we  assign a charge $(-)$ to the site $i$ and a charge $(+)$ 
to the site $j$
(see Fig.~8). Notice that in the path picture one has the same paths as in the
open case but they are now also translated because of the translational
invariance on the ring. Taking into account the action of the generators on
the arches (see Fig.~9b), one can check that the NASEP rules coincide with those
obtained from the Hamiltonian \rf{3.2}. Notice that we have considered  a
quotient of the TLP algebra identifying the picture with  
one non contractible loop and without one. If  $L$ is odd, there are no 
non-contractible loops since one site is attached to the puncture (see 
Fig.~10) and non-contractible loops are blocked by the attachment.
\begin{figure}
\centering
\includegraphics[angle=0,width=0.3\textwidth] {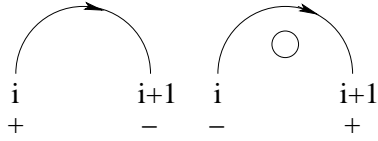}
\caption{
 Assignment of charges to sites connected by an oriented arch
which does
not contain the puncture in the disc and an arch which does contain
the
puncture.
}
\label{prof8}
\end{figure}

 The charged particles presentation of the TLP algebra is as far as we
know new. It is much simpler than the arch presentation. The
particles-vacancies rules follow in a straightforward way.
 A more comprehensive discussion of the periodic Temperley-Lieb 
algebra is going to be presented elsewhere.

 To sum up, if $u = 1$ and half-filling, the NASEP model is conformal
invariant since  this is the case for the Hamiltonian \rf{3.2} (see appendix B). For other
values of $u$ and different densities, this is not necessarily the case.
Moreover one loses integrability and all the informations about the model have
to be obtained from Monte Carlo simulations.

\begin{figure}
\centering
\includegraphics[angle=0,width=0.5\textwidth] {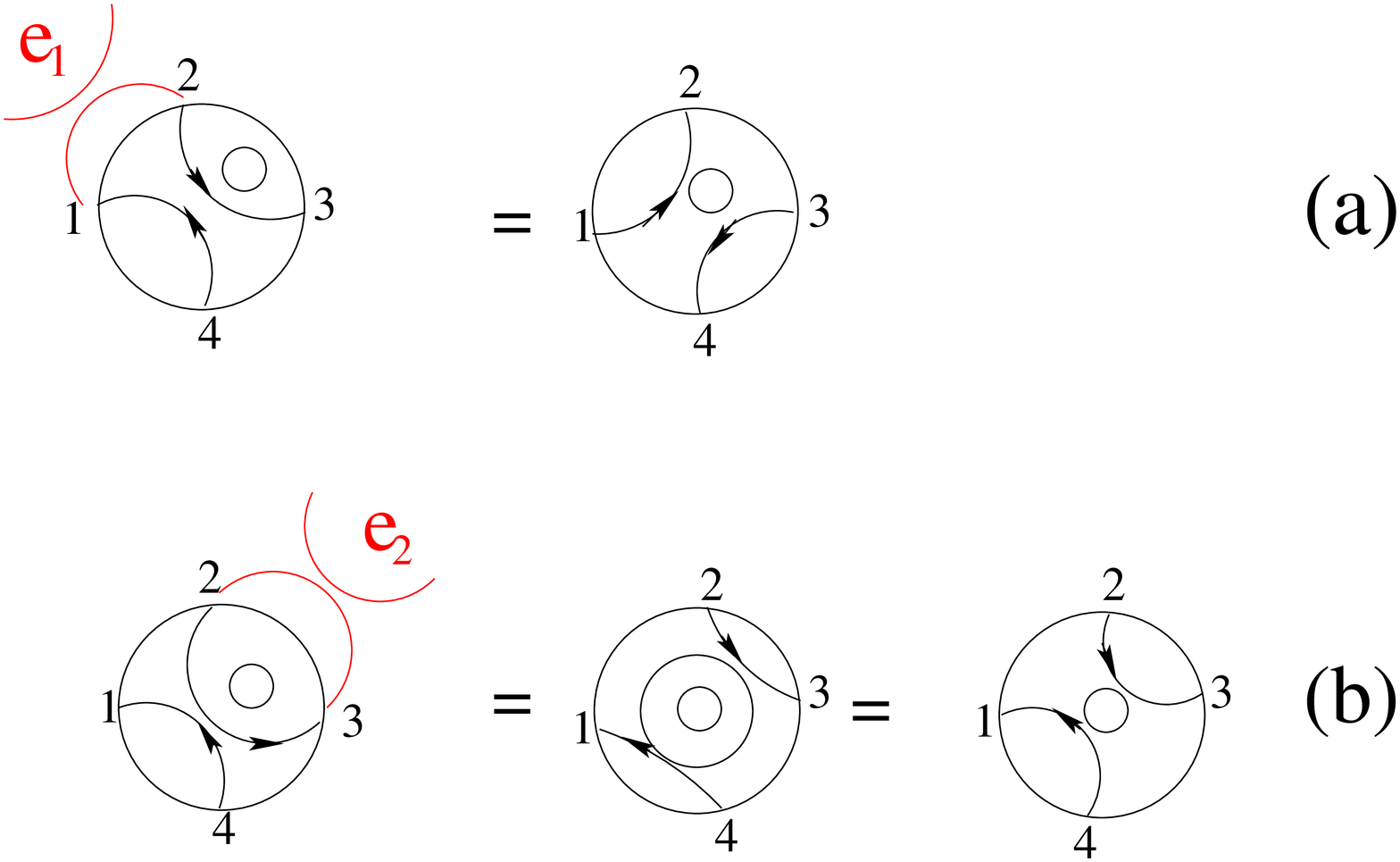}
\caption{
 The action of the generators $e_1$ and $e_2$ of the periodic TLP
algebra for
$L = 4$.}
\label{prof9}
\end{figure}
\begin{figure}
\centering
\includegraphics[angle=0,width=0.2\textwidth] {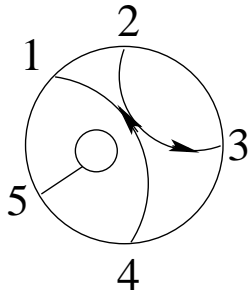}
\caption{
 Loop diagram for $L$ odd, $L=5$ in this case. See the text.
}
\label{prof10}
\end{figure}

 We have shown that the NASEP model on a ring at $u = 1$ and half-filling is
conformal invariant. For $u \neq 1$ and half-filling one can assume that the
phase diagram is the same as in the open system. We remind the reader what
is known in this case. For $u < 1$, the system is gapped, for $u = 1$
it is gapless and conformal invariant (dynamic critical exponent $z = 1$)
and for $u > 1$ it stays gapless with varying critical exponents ($z<1$
decreases if $u$ increases). A study of the correlation functions which is
going to be published elsewhere \cite{JJJ} shows that our assumption is
correct.

 In the next sections we are going to study the behavior of the current in
NASEP. Other features of NASEP are going to be presented in \cite{JJJ}.

\section{Currents in the NASEP model at half-filling in the stationary state}

Like in ASEP, we are interested in the values of the current for various
values of $u$ and densities. Unlike ASEP where the stationary state of the 
periodic system is
trivial while the open system with sources and sinks has relevant physics, 
NASEP
has an interesting phase diagram already for the ring geometry
 In this section we show that in the stationary state currents exist
in NASEP at half-filling and periodic boundary conditions. We expect
their behavior to be dependent on which phase of the model one is. From
the study of the correlation functions \cite{JJJ}, we have found that the phase
diagram for the ring geometry coincides with the one of the open system which
is the RPM. For $0 < u < 1$ one is in a gapped phase, for
$u = 1$ the system is gapless and conformal invariant and for $u > 1$ it is
gapless but not conformal invariant.

If the sites are denoted by $i$ ($i = 1,2,\ldots,L$) with the rules of
Section 2, the
current is defined by the average number  of particles (positive
particles)
which cross the bond $(i,i+1)$ moving from $i$ to $i+1$. By 
convention the current is
positive when the particles move from $i$ towards $i+1$ and negative if
they
move in the other sense (anticlockwise). It is obviously independent on the bond we
choose.

The existence of currents is a novel property of the model and it
should be especially
interesting at $u = 1$. We consider this case first. We should 
expect on dimensional
grounds and
conformal invariance that,  in the large $L$ limit, the current to be 
of the form (1.1), with the sound velocity $v_s=3\sqrt{3}/2$ \cite{CAR} and $C$ 
an universal constant. 

Based on numerical data on lattices up to 18, Pyatov \cite{PYAT} made the
following conjecture for the current ($L$ even)
\be \label{4.2}
J(L) = - \frac{3L}{4(L^2-1)}.
\ee
From this conjecture we get the value $C$ in (1.1).  
The expression \rf{4.2} was checked for
large lattice sizes  using Monte Carlo simulations (see Fig.~11). 
 The existence of a current in the
model was revealed because we have used the particle presentation
of the model. It would have been harder to think of such a quantity in the
link presentation of the TLP algebra. Of course the calculation using
the NASEP version of the model gave the same result. Actually the data
presented in Fig.~11 were obtained using the NASEP version of the model.

\begin{figure}
\centering
\includegraphics[angle=0,width=0.5\textwidth] {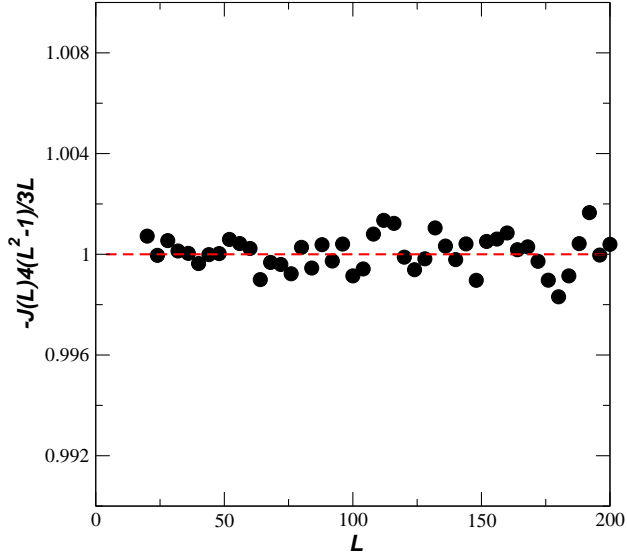}
\caption{
 The current $J(L)$ divided by \rf{4.2} for various values of $L$ at 
$u=1$ and half-filling.  The data are obtained  
from  Monte Carlo simulations of the NASEP model. }
\label{prof11}
\end{figure}
\begin{figure}
\centering
\includegraphics[angle=0,width=0.3\textwidth] {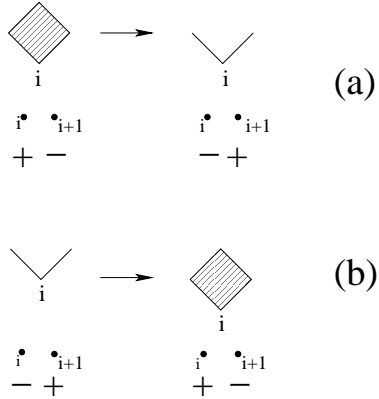}
\caption{
Contributions to the current. a) The desorption of one tile gives a +1 contribution. b) The adsorption of a tile gives a -1 contribution.}
\label{prof12}
\end{figure}

For $L$ odd one obtains $J = 0$ for any size $L$. This result was obtained
from
small lattices and confirmed by Monte Carlo simulations.

One can understand the expression \rf{4.2} in the following way. We
start with the
open system and consider the path and charged particles
presentations. Notice that
desorbing a tile on the site $i$ on the dual lattice gives a positive
contribution
to the current: a positive particle moves to the right (Fig.~12a).
Similarly,
adsorbing a tile on the site $i$ on the dual lattice gives a negative
contribution
to the current (Fig.~12b). Since in the stationary state the average number of
desorbed and adsorbed tiles are the same, the average current is zero.
 This is what
is expected for an open
system.
\begin{figure}
\centering
\includegraphics[angle=0,width=0.6\textwidth] {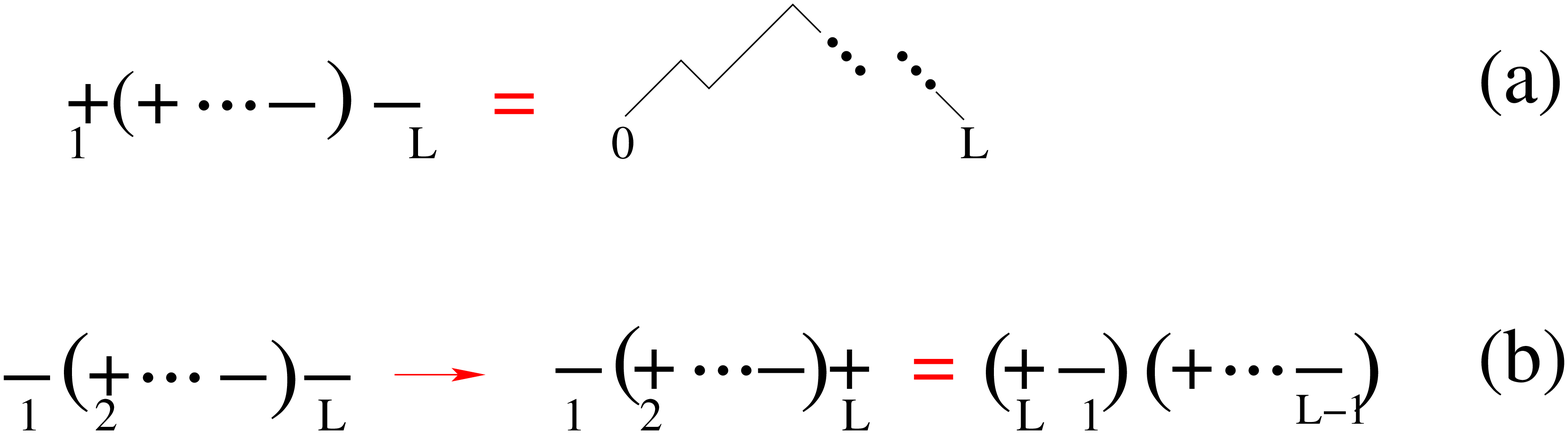}
\caption{
a) A configuration with a cluster having the size of the system. b) 
The effect of a tile hitting the end of the cluster (site $L$).}
\label{prof13}
\end{figure}

 The situation is different if one takes periodic boundary conditions. The
Dyck paths configurations are the same as in the open system but they are
repeated through translations on the circle. The action of the Hamiltonian
on these configurations is similar to the open case with one important
exception. Let us consider a configuration with a single cluster (a Dyck
path which doesn't touch the horizontal axis except at $i = 0$ and $i = L$) 
which
has the size of the system, like in Fig.~13a. 
This is possible only if $L$ is even. We denote this configuration
by $+(+,...,-)-$. Acting on the link $(L,1)$  (see
Fig.~13b) the interchange of the two end particles 
produces the configuration $-(+,\ldots,-)+$ and gives  one negative
contribution to the current. 
 The configuration $-(+,...,-)+$ has $L/2 - 1$ tiles less 
than the configuration $+(+,...,-)-$. These tiles didn't give any positive 
contribution to the current. Since on the average the total numbers of the 
tiles desorbed and adsorbed are equal, one needs to have $L/2 - 1$ adsorbed 
tiles (each with a contribution -1) to compensate for the desorbed tiles. 
 Therefore on the link $(L,1)$ one gets a total contribution to
the current equal to $-1 - (L/2- 1) = - L/2$. On the other hand one can use a conjecture of \cite{JDG}
which gives the probability to have one cluster of size $L$:
\be \label{4.3}
P(L) = \frac{3L}{2(L^2 -1)}.
\ee
Because of translational invariance there are $L$ clusters of size $L$
with a
probability $3/2(L^2 - 1)$ each. Therefore the current is:
\be \label{4.4} 
J(L) = \frac{3}{2(L^2-1)}\left(\frac{-L}{2}\right),
\ee
which coincides with \rf{4.2}. In Appendix A we discuss in detail the
case $L = 4$.
We conclude that the appearance of the current for $L$ even comes 
from transitions taking place at the end of configurations having the size 
of the system. Such configurations, forming a single cluster, do not 
exist for $L$ odd (see Fig.~10), as a consequence the current should be zero.
{
 Using Monte Carlo simulations we have looked at the $L$ 
dependence of the current's fluctuations  in the stationary state:
\be \label{4.4a}
D(L) = <J(L)^2> - <J(L)>^2,
\ee
where $J(L)$ is given by \rf{4.2} and $<J(L)^2>$ is the average of the square 
of the current for a system of size $L$. We have used lattices of size $L = $ 
500, 1000, 2000, 4000, 8000, 16000, 32000 and 64000. A fit to the data 
gave
\be \label{4.4b}
D(L) = a/L^b    \quad {\mbox{  with  }}\quad a = 0.21\pm0.04, \quad b = 0.86\pm0.03.  
\ee
 We are still missing an interpretation of the exponent $b$.

 The time dependence of the current for different values of $L$ was also 
 explored. We used the step initial condition (particles filling one half 
of the lattice and vacancies filling the other half). Using conformal 
invariance, we expect the following behavior of the current:
\be \label{4.4c}
J(t,L) = \frac{1}{t}f(\frac{L}{t}),
\ee
where $f (L/t) \sim  \frac{3}{4}t/L$  for $L/t \to 0$, and 
$J \sim  \frac{3}{4}v_c 1/t$ for $t/L \to  0$  ($v_c 
= 3\sqrt{3}/2)$).        

 The data are shown in Fig.~14 in which we have plotted $t|J(t,L)|$ as a 
function of $L/t$. Lattices of size 300, 600 and 1200 were chosen and the 
time unit was chosen 100 times smaller than the usual one. 
One observes a nice data 
collapse at small values of $L/t$ and large finite-size effects for large 
values of $L/t$ probably due to the choice of the initial condition.
\begin{figure}
\centering
\includegraphics[angle=0,width=0.6\textwidth] {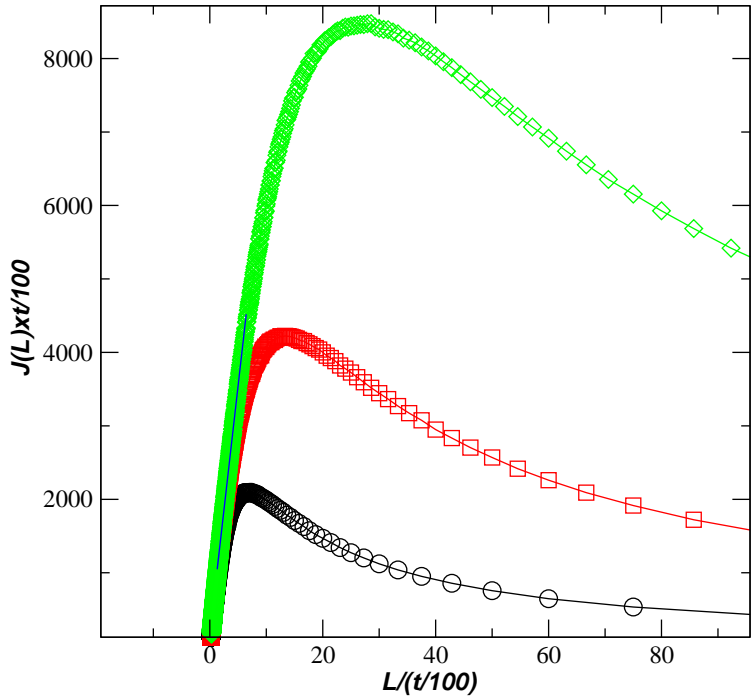}
\caption{
 The current $J(t,L)$ multiplied by $t/100$ as a function of $100L/t$ for the 
sizes $L = 300$ (in green), $L = 600$ (in red) and $L = 1200$ (in black).}
\label{prof14}
\end{figure}

 Taking $u$ away from the value 1 is bound to change the picture. 
For $u< 1$ one has a larger desorption (loss of tiles) and for a given lattice
size $L$, one expects a decrease of the absolute value of the current, the
opposite phenomenon should take place at $u > 1$. At a given value of $u$, 
one
expects the current to vanish exponentially for $u < 1$ (one is in the
gapped phase) and to vanish as a power of $L$ for $u > 1$ (one is in a
non-conformal invariant gapless phase). This is what is observed.

 In Fig.~15 we show the current $J(L)$ as a function of $L$ for $u = 0.75$
obtained in Monte Carlo simulations. A good fit to the data gives:

\be \label{4.5}
J(L) = - 0.0036\exp(-0.0225L),
\ee
confirming the expected exponential decrease of the current with $L$ in
the
gapped phase.

\begin{figure}
\centering
\includegraphics[angle=0,width=0.5\textwidth] {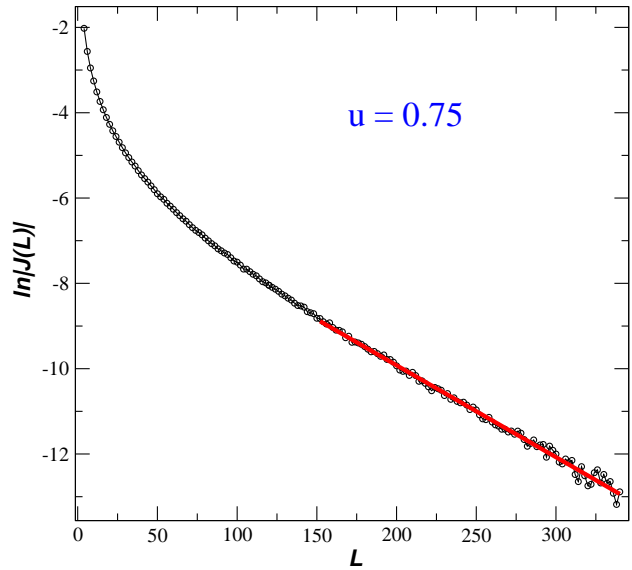}
\caption{
  The current as a function of $L$ ($L$ even) for $u
= 0.75$.
Lattices of size $4, 6, 8,\ldots, 340$ were considered in the Monte Carlo
simulations. The fitting curve is also shown.}
\label{prof15}
\end{figure}

The case $u >1$ is shown in Fig.~16 where the $u = 5$ data are shown. A fit
to the
data gives:
\be \label{4.4aa}
J(L) = -0.917/L^{0.132}
\ee

In the whole phase $u > 1$ the current behaves like $J \sim - 1/L^x$. 
For example $x =0.182$ for $u = 3$. The exponent $x$ decreases 
with $u$, this implies that the
absolute value of the current increases with $u$. This is to be expected
since large clusters have more chance to occur for large values of $u$,
 and hence more clusters
having the size of the system.

\begin{figure}
\centering
\includegraphics[angle=0,width=0.5\textwidth] {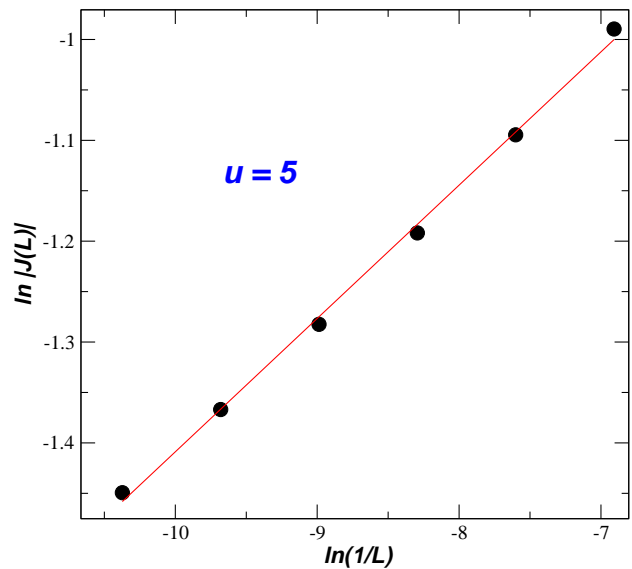}
\caption{
  The current as a function of $L$ ($L$ even) for $u
= 5$ and half-filling, obtained from 
Monte Carlo simulations, for lattice sizes $L$=1000, 2000, 4000, 8000, 
16000 and 32000. The fitting line is also shown.}
\label{prof16}
\end{figure}

 Up to now we have considered the case of $L$ even only and mentioned
that for $u = 1$ the current vanishes for all values of $L$ odd, this is not
the case for other values of $u$. We present now in more detail the data
which show the $u$ dependence of the current for even and odd lattices.
 In Fig.~17 we compare  the behavior of the currents for $L$ even and odd,
for small lattice sizes, for which one can obtain numerically exact results.

%
\begin{figure}[t]
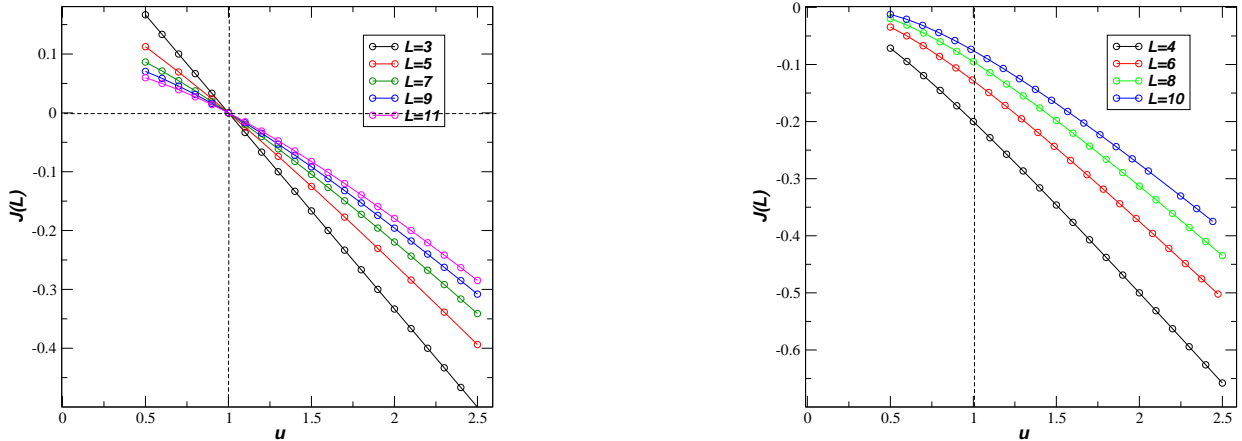

\centering{ \includegraphics [angle=0,scale=0.46]
{fig17a-rpm.eps}\hspace{-3.0cm}\hfill\includegraphics
[angle=0,scale=0.46] {fig17b-rpm.eps} } \caption{
  Exact values of the current for small
lattice
sizes at half-filling: a) even number of sites, b) odd number of sites.}
\label{fig17a}\label{fig17b}
 \end{figure}

For $L$ even, one notices that the current stays negative for all
values of $u$, its absolute value increasing with $u$. For $L$ 
odd the situation 
is different. For $u < 1$ the current is positive, vanishes for $u = 1$ and
changes in sign for $u > 1$. For large lattices and $u > 1$, the ratios of the
 currents even/odd are close to 1. For example for $u=5$ and $L=32000$ we 
obtain: 
$J(L)/J(L+1) = 1.06$.

To sum up, for half-filling in the thermodynamical limit the current
vanishes for all values of $u$. This picture is going to change 
dramatically away from
half-filling.

\section{ Currents at other densities. }

 As we have seen at half-filling, in the thermodynamical limit, the currents
vanish for any value of $u$. This picture changes if the density of particles
(positive particles) $\rho$ is not equal to 1/2. It is convenient to look at
the deficit of density of particles $\eta = 1/2 -\rho$.

 We start by observing that for any $\eta \neq 0$ and any $u$ the system is gapped.
This information comes from the study of correlation functions which have all
an exponential decay \cite{JJJ}.

 At $u = 1$, from an exact analysis of small lattice sizes  and from Monte Carlo
simulations on large lattices sizes, one can conclude that for  any value of $\eta \neq 0$, 
the
currents vanish for large lattice size $L$. For other values of $u$, the current
stays finite. This property is illustrated in Fig.~18 where  the currents are
given for the three values  $\eta = 1/42$, 1/6 and 1/4 and different lattice
sizes. One notices that for a given value of $\eta$ one has data collapse 
of several lattice sizes.
The current is negative for $u > 1$ and positive for $u < 1$.

\begin{figure}
\centering
\includegraphics[angle=0,width=0.5\textwidth] {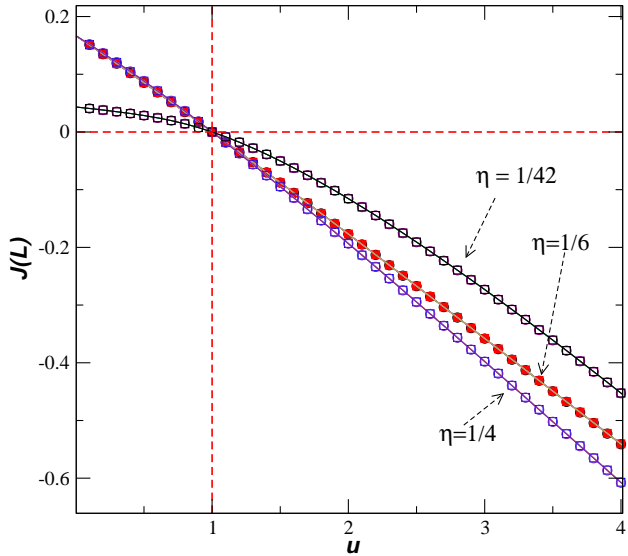}
\caption{
 The currents $J(L)$ as a function
of $u$ for
$\eta= 1/42, 1/6$ and $1/4$, and $L = 1000, 2000, 4000, 8000, 16000$ and
32000.}
\label{prof18}
\end{figure}
The currents approach their asymptotic values with a correction term
$\sim  1/L$.
 This is illustrated in Fig.~19, where we have taken $\eta = 1/4$ and $u
= 5$. A
fit to the data gives $J(L) = -0.72 - 0.82/L$. The same occurs for 
$u <1$. For
example taking $u = 0.75$ and $\eta = 1/4$ one gets $J(L) = 0.04 +
0.05/L$.

\begin{figure}
\centering
\includegraphics[angle=0,width=0.5\textwidth] {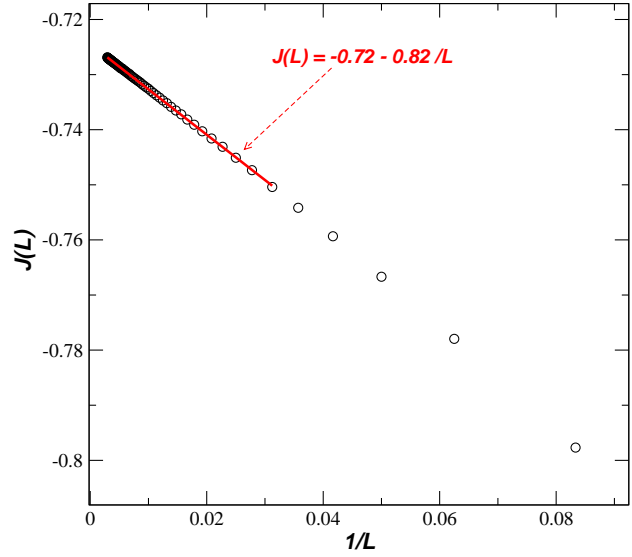}
\caption{
 The current $J(L)$ as a function of $L$ for $u =5$
and $\eta
= 1/4$, $L = 16, 20, 24,...,340$. The fitted curve is also shown.}
\label{prof19}
\end{figure}
 The most interesting feature of the current behavior occurs if, for
fixed $u$,
one looks at its variation with $\eta$. 
A first impression about the
$\eta$ dependence of the current can be obtained from the data for
several values of $u$, shown in Fig.~20, where we considered the small 
lattice size $L =
30$. We observe that, as expected, the currents are
symmetric functions of $\eta$, and that they are positive for
$u < 1$ and negative for $u > 1$. We also see for larger values of $u$, a
kind of plateau around $\eta$ = 0. This observation is a precursor of a
phenomenon seen for large lattices.

\begin{figure}
\centering
\includegraphics[angle=0,width=0.5\textwidth] {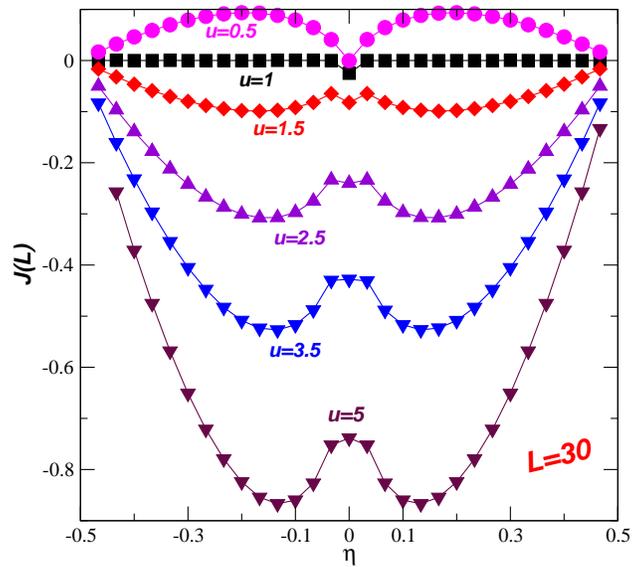}
\caption{
 The current $J(30)$  as
a function
of $\eta$ for several values of $u$.}
\label{prof20}
\end{figure}

 We start by looking  the data for fixed $u = 0.75$ ($u < 1$). They are shown in Fig.~21.
\begin{figure}
\centering
\includegraphics[angle=0,width=0.5\textwidth] {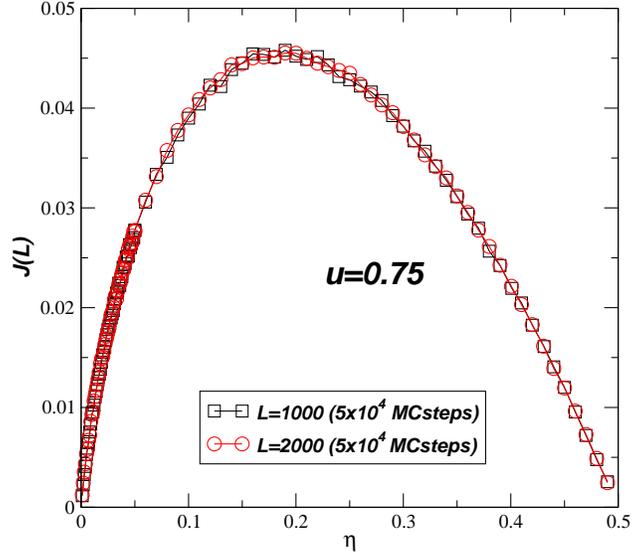}
\caption{
 The current $J(L)$ for $u = 0.75$ as a
function of $\eta$ for lattice sizes $L = 1000$ and 2000. }
\label{prof21}
\end{figure}
 This is the expected behavior since the current has to vanish at $\eta = 0$ 
and 0.5 (there are no particles to carry the current in this latter case). Similar
results are obtained for other values of $u < 1$. 
\begin{figure}
\centering
\includegraphics[angle=0,width=0.5\textwidth] {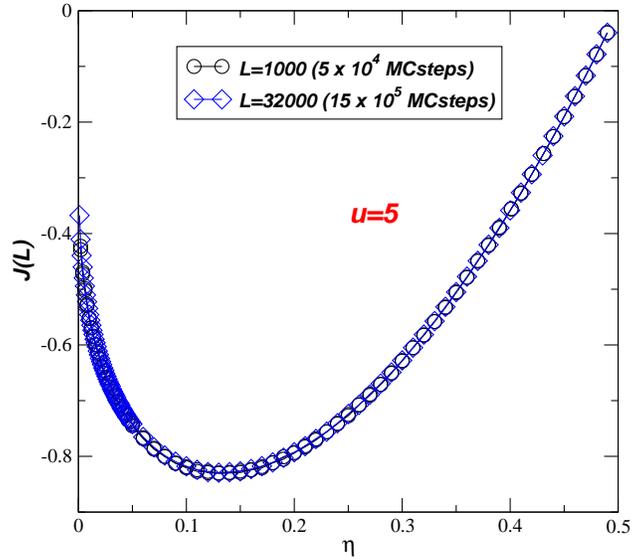}
\caption{
 The current $J(L)$ for 
$u = 5$ as a
function of $\eta$. The data are from  lattice sizes $L = 1000$ and 32000.}
\label{prof22}
\end{figure}
The situation is dramatically
different if $u > 1$. In Fig.~22 we show the data for $u = 5$. 
One notices that
if $\eta$ approaches the zero value (almost half-filling) the current stays
finite. It has the value -0.37 for $\eta = 0.001$ for a lattice of size
$L = 32000$. As we have discussed, the current vanishes at $\eta = 0$ and this
implies a phase transition of a new kind since the current 
has a discontinuity. 
 The discontinuity increases with $u$ (it vanishes at $u \leq 1$).
This is illustrated in Fig.~23 where we show, as a function of $u$,  the values of $J(L)$ for 
$\eta = 0.001$  
 and three lattice sizes.

The appearance of this phase transition came  as a surprise. We have tried to
understand its origin by looking at another quantity which is relevant to the
existence of the current: the density of vacancy-particle pairs $\rho_{v-p}$
 (negative
positive particles pairs or valleys in the language of Dyck paths) in the
stationary states. In the case $\eta = 0$, this quantity was already
studied for various values of $u$ for the open system and we don't expect big
changes for the periodic system. This is at least the situation for $u = 1$ for
which this density is equal to $\rho_{v-p}=3/8 = 0.375$ for the open and  periodic
systems\cite{MNG}. We expect this quantity to decrease for larger values of $u$ \cite{CAR}.
\begin{figure}
\centering
\includegraphics[angle=0,width=0.5\textwidth] {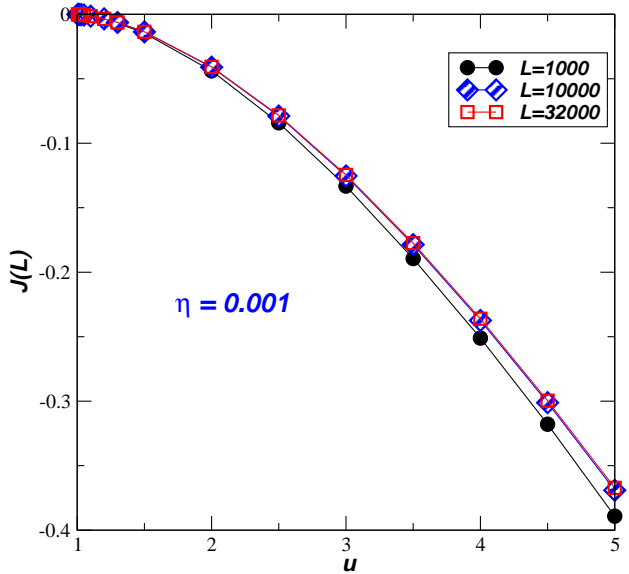}
\caption{
Illustration of the discontinuity of the current for $\eta \neq 0$. The current as a function of $u$ for $\eta=0.001$ for the lattice sizes $L$=1000, 10000 and 
 32000.}
\label{prof23}
\end{figure}
\begin{figure}
\centering
\includegraphics[angle=0,width=0.5\textwidth] {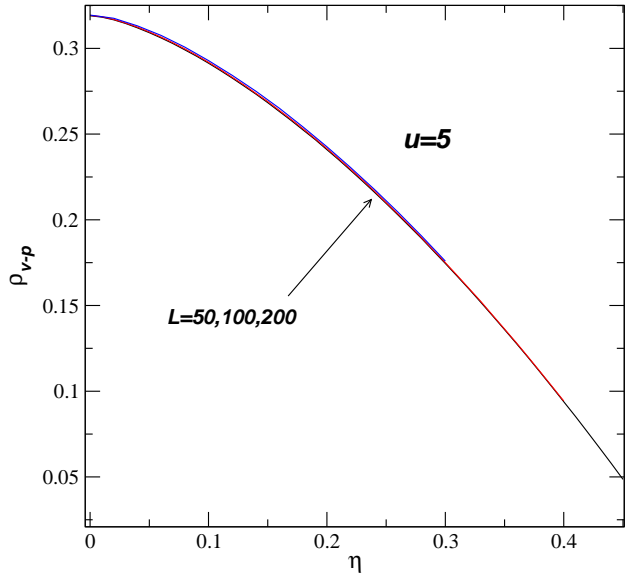}
\caption{
 The density of vacancy-particle pairs
as a function of $\eta$ for $u = 5$ and  lattice sizes $L = 50, 100$ and 200.}
\label{prof24}
\end{figure}

 In Fig.~24 we show for $u = 5$ the density of  pairs 
$\rho_{v-p}$ as a
function of $\eta$ for three lattice sizes. One clearly sees data collapse
already for relatively small lattice sizes, and no discontinuity is observed. Moreover the value 0.31, at  the density
at $\eta$ = 0 is compatible with the known result \cite{CAR} for the open
system. 
 This observation made us think that may be 
the phase transition is a red herring.

 If we do have a novel phase  transition we expect the current to have 
the following behavior at finite $L$:
\be \label{aa}
J(L) = A \eta^x,
\ee
with the exponent $x$ decreasing with $L$ such that in the thermodynamic limit, 
$x = 0$. If on the other hand we don't have the novel phase transition, one 
can explain the data by having just very small values of the exponent $x$ which 
fake the phase transition. If this is the case, $x$ could even slightly increase
with $L$. Taking into account that in Figs.~21 and 22 we have used lattices 
up to 32000 sites and went to densities very closed to the half-filling 
value ($\eta = 0.001$), in order to clarify the issue, we decided to 
consider lattices up to 256,000 sites and values of $\eta$ as small as
$0.000005$. In order to interpret the data we have to keep in mind that we 
are in a gapped phase and, consequently, at a fixed value of $\eta$ the 
data should converge exponentially for large values of $L$. 
   
 In Fig.~25 for $u = 3$, we show the current as a function of $\eta$ for 
several lattice sizes between 500 and 256,000. One notices that for very small 
fixed values of $\eta$ the absolute value of the current keeps 
slightly decreasing even for very large lattices. The existence of the novel
phase transition, would have implied an $L$ independent constant value of 
the current.
\begin{figure}
\centering
\includegraphics[angle=0,width=0.5\textwidth] {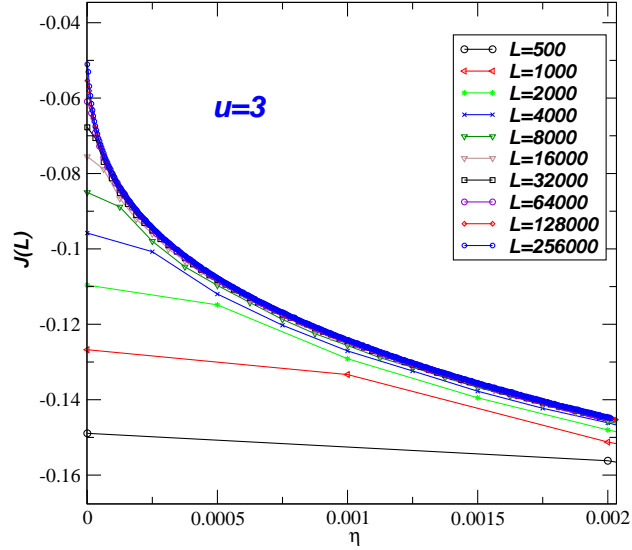}
\caption{
The current as a function of $\eta$ for $u =3$. Lattices with sizes 
increasing by a factor of 2 from 500 to 256.000 were considered.}
\label{prof25}
\end{figure}
 The situation becomes even clearer if we examine the data shown in the 
next figure (Fig.~26) where the results for the largest lattices 
($L = 128,000$ and $L = 256,000$) are presented only. The results of the fits 
using \rf{aa} are very interesting. One obtains:
\ba \label {bb}
&& A = 0.333, \quad x = 0.152, \quad     L = 128,000  \quad   (\mbox{region 1}),
 \nonumber \\
&&A = 0.338, \quad x = 0.156, \quad     L = 256,000 \quad     (\mbox{region 1}), 
\nonumber \\
&&A = 0.293, \quad x = 0.142, \quad     L = 256,000 \quad    (\mbox{region 2}).
\nonumber
\ea 
The region 1 contains "larger" values of $\eta$, Region 2 "smaller" 
values. Within errors, the values of $x$ for the two regions coincide 
and are not equal to zero. 
The difference between the values of $x$ for the two Regions is minimal. A 
good fit for all the data gives $A = - 0.30\pm 0.05$, $x = 0.145 \pm 0.005$. 
We 
conclude that there is no novel phase transition. The current vanishes 
smoothly albeit in a non-analytic way when $\eta$ vanishes. This could 
have been expected since we have a transition from a gapped phase to a 
gapless one but it was a long way to get to this conclusion. 
\begin{figure}
\centering
\includegraphics[angle=0,width=0.5\textwidth] {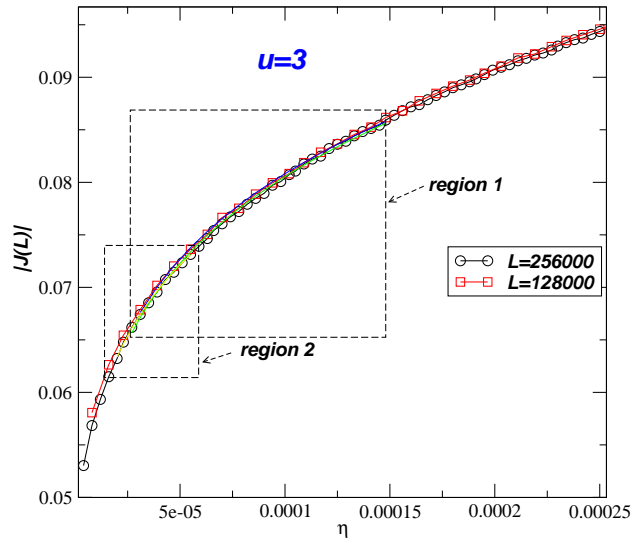}
\caption{
The very small $\eta$ domain. The absolute value of the current as a 
function of $\eta$ for $L = 128,000$ and 256,000.}
\label{prof26}
\end{figure}

 The existence of a non-analytic behavior of the current, but albeit no 
phase transition of a new kind, was seen for various values of $u$. The 
exponent $x$ and the factor $A$ are given for several values of $u$ in table 1. It looks 
like $x$ increases to the value one if $u$ approaches the value 1. 
This is a 
reasonable guess since $x = 1$ for $u < 1$ (we have no phase transition in 
this domain and we expect an analytic dependence on $\eta$). In order to 
confirm the value $x = 1$ for $u < 1$, we have included several values of 
$u < 1$ in
table 1. For all these values of $\eta$ one finds $x\approx 1$ indeed.
\begin{table}
\centerline{
\begin{tabular}{|c|c|c|}
  \hline
   $u$       & $A$  &    $x$  \\
\hline
0.0625 & $1.90 \pm 0.01$ & 1 (*) \\
0.125 & $1.85 \pm 0.01$ & 1 (*) \\
0.25 & $1.70 \pm 0.01$ & 1 (*) \\
0.5 & $1.40\pm 0.05$ & $0.994 \pm 0.005$ \\
0.75 & $1.13\pm0.05$ & $0.99\pm0.01$ \\
1 & 0.0  & - \\
1.25 & $-0.043\pm0.001$ & $0.343\pm0.005$ \\
1.5 & $-0.060\pm0.001$ & $0.245\pm0.002$ \\
1.75 &$-0.089\pm0.001$ & $0.205\pm0.005$ \\
2   & $-0.130\pm 0.005$ & $0.180\pm0.005$ \\
3  & $-0.30\pm0.05$ & $0.145\pm0.005$ \\
4  & $-0.48\pm0.02$ & $0.12\pm0.01$ \\
5  & $-0.70\pm0.05$ & $0.110\pm0.005$ \\
7  & $-1.15 \pm0.05$ & $0.085\pm0.05$ \\
10 & $-1.75\pm0.05$ &$0.065\pm0.05 $ \\
  \hline
\end{tabular}}
 \caption{
 Values of the parameters $A$ and $x$ in the behaviour \rf{aa}  of the
density of current $J$ as a function of the density $\eta$, for several values of $u$. These estimates were obtained by considering values of $\eta <0.00025$ and lattice sizes $L=128K$ and $L=256K$ for the periodic RPM model.
 The lines with (*) are obtained by a linear fit, since in this case the fit is much better.}
\label{table1}
\end{table}


\section{Conclusions}

  In this paper the Raise and Peel model  is reformulated as a 
nonlocal 
asymmetric exclusion process (NASEP). 
We extend and study the model in  the 
case of periodic boundary conditions and arbitrary densities of particles. 
   NASEP depends on two parameters 
$u$
and the density of particles $\rho$. The parameter $u$ gives the
forward-backward asymmetry of the model.

  At half-filling and $u = 1$, the system is conformal invariant (dynamic 
critical exponent $z = 1$) and the spectrum is known in the finite-size scaling 
limit \cite{DGFE}. The system is integrable (see Appendix B) and the probability 
distribution function describing the stationary state has remakable 
combinatorial properties \cite{RAST,CAS}. Still at half-filling, if $u < 1$, the system is 
gapped while for  $u > 1$, the system is gapless with the critical exponent $z$ 
 varying continuously with  $u$ ($z(u) <1$). The function $z(u)$ decreases with $u$. 
At any density $\rho\neq 1/2$ 
 the system is gapped. This implies that getting $\rho$ closed to the 
value 1/2, we have no phase transition if $u < 1$, a usual phase transition 
if $u = 1$ and possibly a new kind of phase transition if $u > 1$.

  We have studied the current in NASEP. This was possible because of the 
extension of the model to periodic boundary conditions. In stationary 
(nonequilibrium) states the current can be seen as an order parameter and 
its properties should reflect the phase diagram. This is indeed the case.
  At half-filling and even lattice size $L$, the current vanishes exponentially for $u < 1$, 
and as $L^{-x(u)}$ otherwise ($x(u)<1$). If $u = 1$ it has the 
expression (1.1) with $C$ an universal constant. This implies that at 
half-filling the current vanishes in the thermodynamic limit for any $u$. 
For $L$ odd the current vanishes identically for any $L$ and $u$. If we are 
not at half-filling, the current stays finite in the thermodynamic limit for
any density and asymmetry $u$.

  It is interesting to see how the current vanishes as a function of $\eta 
= \rho - 1/2$ when $\eta$ approaches the value zero. For $u < 1$, the 
current vanishes 
 linearly with $\eta$, for $u = 1$ it vanishes for any number of 
sites. For $u > 1$ it vanishes like $\eta^x$ where the exponent $x$ 
decreases with $u$, getting very small values for moderate values of $u$. 
Finding the exponent $x$ using Monte Carlo simulations (see table 1) was not 
an easy task. One had to use very large lattices (up to 256,000 sites).  

  This paper is going to be followed by a sequel \cite{JJJ} in which we present 
the fluctuations of the current and various correlation functions.

   In Section 3 we  derive  the model for $u = 1$ 
and half-filling using the periodic Temperley-Lieb algebra. In Appendix B 
we make the connection of the model with integrable quantum chains and 
derive the expression of the spin current.  
  The 
fact that the spin current vanishes for any size $L$ ($L$ odd) is also shown.
 Notice that the spin current has a behavior similar to the NASEP current. 

  The reader might have noticed that the expression Bethe ansatz was not 
used in the text except in Appendix B. This is not an accident. Unlike 
in TASEP or PushASEP where a lot of work was done 
(see \cite{FRF,PHS,corwin,gueudre} and references 
therein), up to now a whole class of questions were not asked yet in the 
case of NASEP. For example, we didn't look at large time fluctuations of 
observables by starting with flat or step initial conditions \cite{BFE,BFS} and 
checked for the existence of an equivalent of so-called Airy processes. 
Since for $u = 1$ and half-filling the system is integrable one might hope 
that some pretty properties might show up.

  We have to mention that for an open system the description 
of NASEP can be found in Section 2. The formulation of the model in the 
presence of sources and sinks at the boundaries remains to be done. 
Finally, a very relevant question about out work stays still without an answer: 
we keep looking for physical applications. 

\section{Acknowledgments} 

 VR would like to thank DFG (project RI 
31716-1) and FCA to FAPESP and CNPq (brazilian agencies) for financial support. We also thank V. Priezzhev for very fruitful 
discussions. 

\appendix
\section {Appendix:  The current for four and three particles on a ring for $u=
1$ }

 We present first the calculation of the current on a ring in the case of two 
$(+)$ and two $(-)$ ($L$ even) particles and next the case of two $(+)$ and 
one $(-)$ particles ($L$ odd). This calculation will make clear why one has a current in the first case and not in the second.

  There are 6 configurations for $L = 4$ shown in Fig.~7. 
We consider the configuration $++--$ on the sites 1,2,3 and 4. 
In this configuration one has a tile on top of the substrate. 
We apply the rules of Section 2 (see (2.1)-(2.4))
 on each of the four bonds in order to find out which configurations one obtains. We keep track on the number of 
tiles lost or won and of the current on the bond. We repeat the same procedure also for the
 configuration $+-+-$  on the ordered 4 sites. No tiles are present in this case. The results are shown in table 2 where we denote by $[\ \ ,\ \ ]$ a bond.
\begin{table}[t]
\centerline{
\begin{tabular}{|c|c|c|}
  \hline
 $|{\mbox{in}}> \rightarrow |{\mbox{out}}>$ & Tiles & Current \\ \hline 
$[+_1 +_2] -_3 -_4 \rightarrow +_1 -_2 +_3 -_4$ & -1 & +1 \\
$+_1 +_2 [-_3 -_4] \rightarrow +_1 -_2 +_3 -_4$ & -1 & +1 \\
$+_4] +_1 -_2 [-_3 \rightarrow +_4 -_1 +_2 -_3$ & -1 & -1 !!! \\
$+_1 [+_2 -_3] -_4 \rightarrow +_1 +_2 -_3 -_4$ & 0 & 0 \\
$[+_1 -_2] +_3 -_4 \rightarrow +_1 -_2 +_3 -_4$ & 0 & 0 \\
$+_1 -_2 [+_3 -_4] \rightarrow +_1 -_2 +_3 -_4$ & 0 & 0 \\
$+_1[ -_2 +_3 ]-_4 \rightarrow +_1 +_2 -_3 -_4$ & +1 & -1  \\
$+_1] -_2 +_3 [-_4 \rightarrow +_1 +_2 -_3 -_4$ & +1 & -1  \\
\hline
\end{tabular}
} \caption{
	The dynamics of 4 sites. The other processes not shown are 
obtained by cyclic permutation of the presented ones.}
\label{table2}
\end{table}
The results for other configurations are obtained by simple permutations. 
One can diagonalize the Hamiltonian obtained from table 2 and find that 
in the stationary state each of the four configurations with two adjacent
 (+) charges have the
probability 1/10 and each of the two configurations with no tiles 
have a probability 3/10.
 
 Let us first note that in the stationary state (see table 2), the average number of 
tiles desorbed equal to $3\times 4\times 1/10$ is equal to the number 
of tiles adsorbed $2\times 2 \times 3/10$. With one exception, 
each desorbed tile contributes one positive unit to the current while 
each adsorbed tile gives a negative unit. 
If one wouldn't have an exception, the current would have been zero like in the open system.
 The exception occurs when one considers the bond $[-\; +]$ on the sites
 $[4,1]$. Although one looses a tile one gets a negative contribution to 
the current. This phenomena is the origin of a negative current. A simple 
arithmetic gives a current equal to -1/5. There is another simpler 
derivation of the value of the current. If on the bond $[4,1]$ one would 
have had a (+1) contribution to the current, the total current would be 
equal to zero (like the balance of the number of tiles). One has 
therefore subtract and add this value to obtain a net contribution of -2 
to the current. Since the probability of the configuration is 1/10, 
one recovers the value -1/5 in agreement with (4.2).

 This simple way of reasoning does not apply for an odd number of sites. 
Let us consider the case $L = 3$ on a ring as an example. One has 3 
configurations with a probability 1/3 each: $++-$, $+-+$ and $-++$. 
For an open system one has 2 of them in the relevant subspace: $++-$ and 
$+-+$ and one can attribute a tile to the first configuration and none to 
the second. One can then show that the average number of tiles in the 
stationary state is 1/2. In the periodic system the 3 configurations can 
be seen as having all one tile or none. Applying the rules (2.1) one can show that the current vanishes.

\section{Appendix: The spin current in the spin presentation of the periodic 
Temperley-Lieb algebra}

 The time evolution in NASEP is given by a non-hermitian Hamiltonian. 
We are going to show that his Hamiltonian also acts taking  a different 
basis, in a sector of a hermitian Hamiltonian that we are going to derive. 
The new 
Hamiltonian describes an integrable quantum spin chain about which a lot is 
known. We will compute the spin current in this chain and compare it with 
the current derived in Section 4. We have to stress that all results 
presented in this Appendix are related to the case $u = 1$ and half-filling 
of NASEP.

 The TLP algebra at the semigroup point is defined in Eq.~(3.3). For $L$  
odd, the generators have the following presentation in terms of Pauli 
matrices \cite{MNG}:
\be \label{B.1}
e_i = \sigma_i^+\sigma_{i+1}^- +\sigma_i^-\sigma_{i+1}^+ - \frac{1}{4} 
\sigma_i^z\sigma_{i+1}^z -i\frac{\sqrt{3}}{4}(\sigma_i^z -
\sigma_{i+1}^z) +\frac{1}{4},
\ee
where $i = 1,2,...L$, and  $e_{L+1} = e_1$. 
Using (3.2) one obtains the Hamiltonian 
in the spin representation:
\be \label{B.2}
H= -\sum_{i=1}^L\left[ 
\sigma_i^+\sigma_{i+1}^- +\sigma_i^-\sigma_{i+1}^+ - \frac{1}{4}\sigma_i^z\sigma_{i+1}^z -\frac{3}{4}\right], \quad \mbox{($L$ odd)}.
\ee
This is a hermitian periodic Hamiltonian. The picture is different if $L$ 
is even. The first $L - 1$ generators have the expression \rf{B.1} but 
$e_L$ is different:
\be \label{B.3}
e_L = \sigma_L^+\sigma_1^- e^{i\phi} + \sigma_L^-\sigma_1^+e^{-i\phi} 
-\frac{1}{4}\sigma_L^z\sigma_1^z -i\frac{\sqrt{3}}{4} (\sigma_L^z - \sigma_1^z)
+\frac{1}{4},
\ee
where $\phi = -2\pi/3$. Using (3.2) one obtains the Hamiltonian:
\ba \label{B.4}
H&=& -\sum_{i=1}^{L-1}\left[ 
\sigma_i^+\sigma_{i+1}^- +\sigma_i^-\sigma_{i+1}^+ + \frac{1}{4}\sigma_i^z\sigma_{i+1}^z -\frac{3}{4}\right] \nonumber \\ 
&&-\sigma_L^+\sigma_1^- e^{i\phi} 
- \sigma_L^-\sigma_1^+e^{-i\phi} 
+\frac{1}{4} \sigma_L^z\sigma_1^z +\frac{3}{4},
\quad \mbox{($L$ even)}.
\ea
 This is an hermitian Hamiltonian with 
twisted boundary condition: $\sigma_{L+1}^{\pm} = \exp(\mp i\phi) 
\sigma_1^{\pm}$, characterized by the twist angle $\phi$.  It is known 
that the two Hamiltonians \rf{B.2} and \rf{B.4} are integrable and their 
finite-size scaling spectra are known \cite{BGR}. The operator
\be \label{B.5}
S^z = \sum_{i=1}^L \sigma_i^z
\ee
commutes with the Hamiltonians and for half-filling ($L$ even) 
the spectrum 
of NASEP coincides with the spectrum of the Hamiltonian \rf{B.4} in the 
sector $S^z = 0$. For $L$ odd NASEP is in the $S^z = 1/2$
 (or equivalently $-1/2$) sector. 
The basis of positive and negative particles (particles 
and vacancies) is however different from the spin up $\leftrightarrow$ down 
spin  basis of Pauli matrices. There is a similarity transformation which 
relates the two basis. Following \cite{SBS} we are going to compute the 
spin currents for the $L$  odd and even and compare the obtained currents 
with those of NASEP. To do so, using a similarity transformation we bring the 
Hamiltonian   \rf{B.4} to the form:
\be \label{B.6}
H= -\sum_{i=1}^L\left[
\sigma_i^+\sigma_{i+1}^- e^{i\frac{\phi}{L}} 
+\sigma_i^-\sigma_{i+1}^+e^{-i\frac{\phi}{L}}  
- \frac{1}{4}\sigma_i^z\sigma_{i+1}^z -\frac{3}{4}\right] \quad (L {\mbox{ even}}).
\ee
 The spin current operator on the bond $[i,i+1]$ is
\be \label{B.7}
J_i^z = i(\sigma_i^+\sigma_{i+1}^- - \sigma_i^-\sigma_{i+1}^+).
\ee 
 If $E(\phi, L)$ is the ground-state energy for system of size $L$ and 
 twist angle 
$\phi$. Using \rf{B.6}, \rf{B.7} and translational invariance, 
one obtains in 
leading order in $L$ the following expression for the average value of the 
spin current:
\be \label{B.8}
J^z = \bra{0}J_i^z\ket{0} = -\frac{\partial E(\phi,L)}{\partial \phi}.
\ee
 Since for $L$ odd there is no twist (one has periodic boundary 
conditions), it follows that the current vanishes, like in NASEP. For $L$ 
even one can use the results of Ref.~\cite{ABB} (Eq.~(3.25)):
\be \label{B.9}
\frac {\partial E(\phi,L)} {\partial \phi} = \frac {3 v_s \phi}{4 \pi L},
\ee
where $v_s = 3\sqrt{3}/2$ is the sound velocity. Taking into account that
$\phi = - 2\pi/3$ one obtains finally
\be \label{B.10}
J^z = -\frac{3\sqrt{3}}{4 L},
\ee
which, up to a factor $\sqrt{3}$, coincides with NASEP current (4.4) for 
large 
values of $L$. Notice that we did a quantum mechanical calculation in an 
equilibrium state and did not consider  the stationary state of a 
stochastic process.

\end{document}